\documentclass[11pt]{article}
\usepackage{appb}
\usepackage{epsfig}

\begin{document}

\title{The $(\pi^-,\gamma\gamma)$ program at TRIUMF }
\author{ Piotr A. \.Zo{\l}nierczuk\thanks{e-mail:zolnie@pa.uky.edu}\\
on behalf of the RMC collaboration\\
\address{ 
Dept. of Physics and Astronomy, University of Kentucky\\ 
Lexington, KY 40506, USA}}

\maketitle
\begin{abstract}
We report the first observation of the doubly-radiative
decay mode of pionic hydrogen 
($\pi^- p \rightarrow \gamma \gamma n $) and deuterium
($\pi^- d \rightarrow \gamma \gamma X $)
using the RMC pair spectrometer at TRIUMF~\cite{Wright:92}.
The process is interesting in the context of 
the $\pi$-Compton scattering~\cite{Zolnierczuk:98}
and the $\chi$PT predictions of the pion polarizability~\cite{Holstein:90}. 
We present our preliminary values for the
B.R.$_{[\pi^-p\rightarrow\gamma\gamma n]}=3.8\times10^{-5}$ and 
B.R.$_{[\pi^-d\rightarrow\gamma\gamma X]}=1.6\times10^{-5}$.
Our $\pi^-p$ data indicates the dominance of $\pi\pi\rightarrow\gamma\gamma$
mechanism. The $\pi^-d$ data shows no evidence for the $d^*_1$ dibaryon.
\end{abstract}

\section{Motivation}
The double radiative decay of pionic atoms $\pi^-A\rightarrow\gamma\gamma X $ 
was investigated theoretically
by Ericson and Wilkin~\cite{Ericson:75} a quarter of a century ago.
They predicted that the dominating reaction mechanism was the annihilation
of the stopped, real $\pi^-$ with a soft, virtual $\pi^+$, {\em i.e.} 
$\pi\pi\rightarrow\gamma\gamma$. As they advertised, the annihilation mechanism
affords a selective coupling to soft pions, and special sensitivity to renormalization
effects in nuclear matter. 

\begin{figure}[htbp]
    \leavevmode
    \begin{center}
    \begin{tabular}{lll}
	a) & b) & c) \\
      \epsfig{width =3cm,angle=90,figure=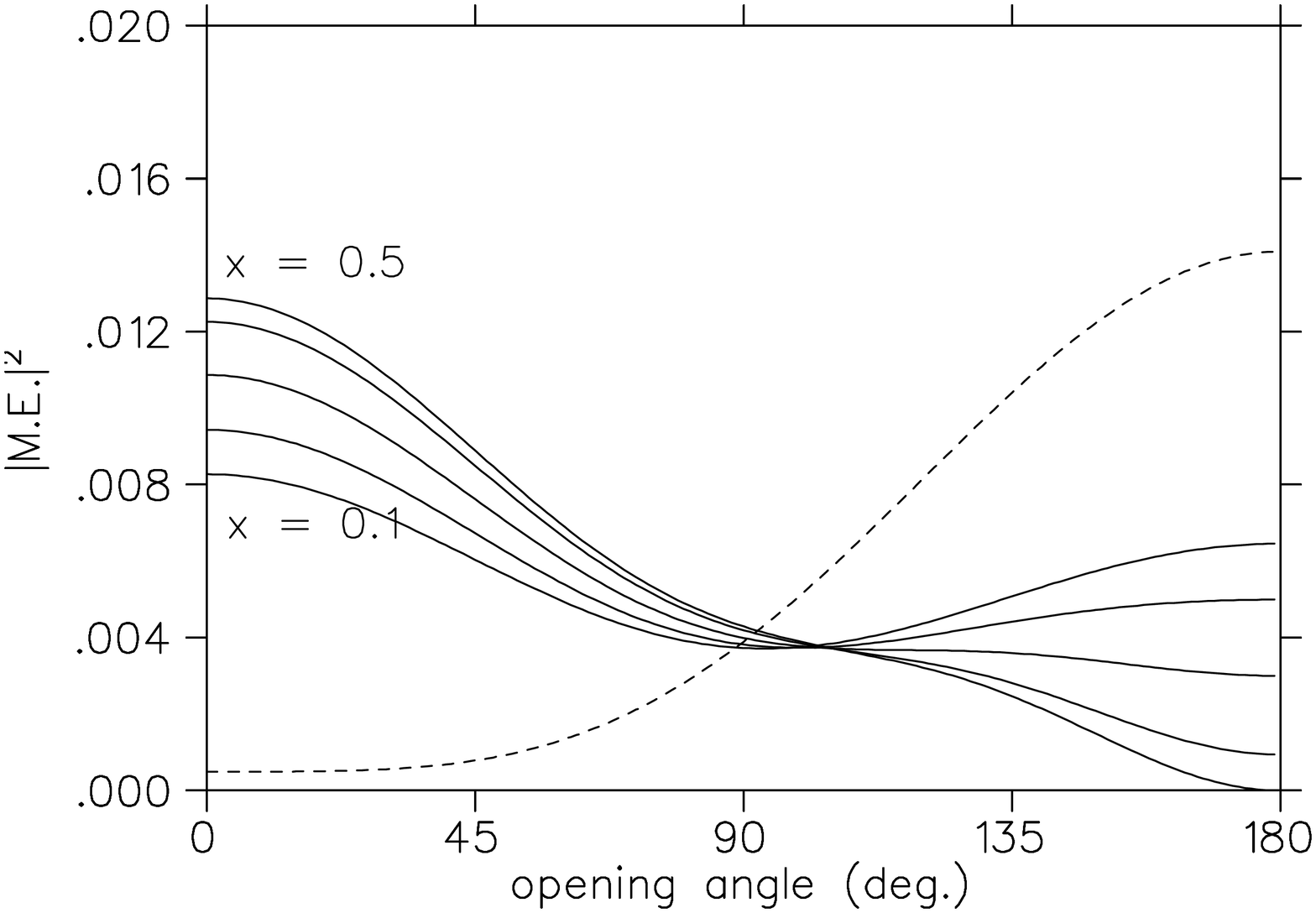}     &
      \epsfig{height=3cm,width=3cm,figure=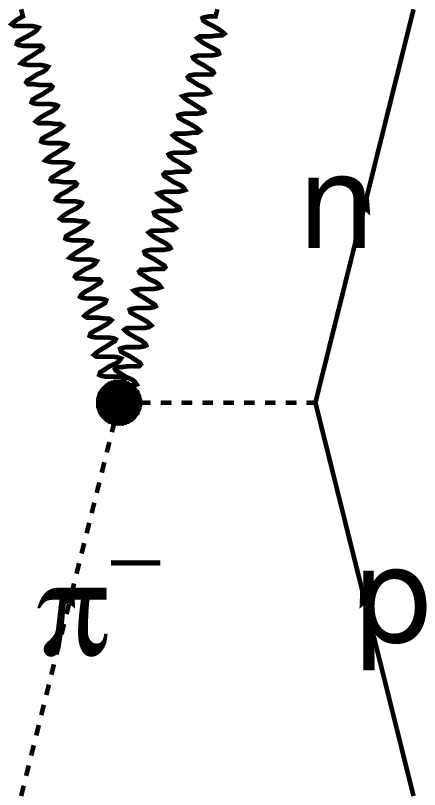} &
      \epsfig{height=3cm,width=3cm,figure=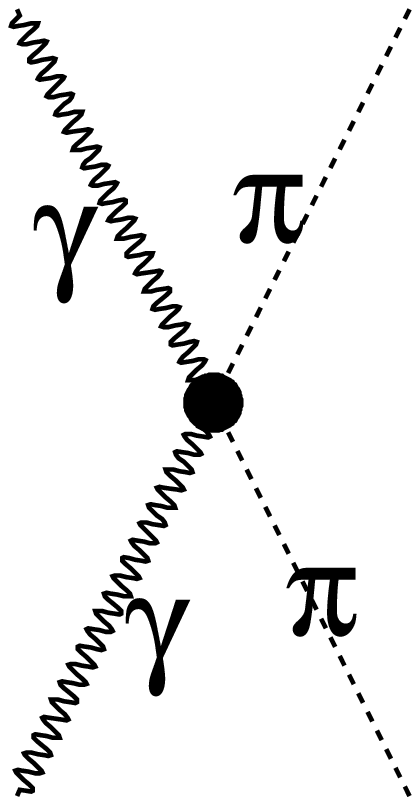}  
    \end{tabular}
    \caption{
      a) Beder predictions for $\pi^-p\rightarrow\gamma\gamma n$ amplitude.
      Solid lines represent the contribution from the annihilation graph for various 
      photon energy sharing parameters $x=\frac{E_1}{E_1+E_2}$. The dashed one 
      is the bremsstrahlung contribution.
      b) $\pi^-p\rightarrow\gamma\gamma n$ annihilation graph,
      c) pion Compton scattering graph.}
    \label{fig:rmc_theory}
    \end{center}
\end{figure}

In order to understand the $\gamma\gamma$ decay mode of a pionic atom,
it is necessary to understand the elementary process of 
$\pi^-p\rightarrow\gamma\gamma n$.
Several authors have made tree--level calculations of the $\gamma\gamma n$ capture mode
of pionic hydrogen. The most recent published calculations are due to 
Beder~\cite{Beder:79},
who obtained 5.1$\times$10$^{-5}$ for the $\gamma\gamma n$ branching ratio.
Beder also pointed out the importance of the pion annihilation diagram,
especially at small photon opening angles (see Fig.~\ref{fig:rmc_theory}a).

The first attempt to observe the two photon emission in hydrogen pion capture
was made by Vasilevsky~et.al.\cite{Vasilevsky:69} at JINR Dubna. 
They obtained an upper limit of the $\gamma\gamma n$ branching ratio, 
B.R. $\le 5.5\times10^{-4}$, i.e. a value ten times larger than 
the theoretical prediction of Ref.~\cite{Beder:79}.

The predicted dominance of the $\pi\pi\rightarrow\gamma\gamma$
annihilation amplitude in $\pi^-p\rightarrow\gamma\gamma n$ allows 
to study pion Compton scattering~\cite{Zolnierczuk:98}.
The Feynman diagram in Fig.\ref{fig:rmc_theory}b 
can be viewed as the annihilation of a real pion with a virtual pion
$\pi^-\pi^+\rightarrow\gamma\gamma$ or, via crossing symmetry,
as the transition of a real pion to a virtual pion via Compton scattering
$\gamma\pi\rightarrow\gamma\pi$ (Fig.\ref{fig:rmc_theory}c).

Pion Compton scattering is a probe of the pion polarizability
$\alpha_E^{\pi^\pm}$ (see e.g.\cite{Holstein:90}).
A summary of the current determinations  of the pion polarizability
is shown in Tab.~\ref{tab:rmc_pionpol}.
\begin{table}[htbp]
  \begin{center}
    \leavevmode
    \caption{Experimental determinations of pion electric polarizability $\alpha_E^{\pi^\pm}$.}
    \begin{tabular}{llll}
      \hline
      Reaction & $\alpha_E^{\pi^\pm}\; (\times 10^{-4}\;fm^3)$ & Experiment & Reference\\
       \hline
      $\pi A \rightarrow \gamma \pi A $    & $6.8 \pm 1.4 \pm 1.2 $  & 
	Serpukhov & \cite{Antipov:84}\\ 

      $\pi p \rightarrow \gamma \pi p $    & $20 \pm 12 $            & 
	Lebedev & \cite{Aibergenov:86}\\

      $\gamma \gamma \rightarrow \pi \pi $ & $19.1 \pm 4.9 \pm 5.6 $ & 
	PLUTO & \cite{Babusci:92}\\

      $\gamma \gamma \rightarrow \pi \pi $ & $2.2 \pm 1.6 $          & 
	MARK II &\cite{Babusci:92}\\

      \hline
    \end{tabular}
    \label{tab:rmc_pionpol}
  \end{center}
\end{table}
The extracted pion polarizabilities have
large uncertainties and are (except MARK II)
substantially larger than the $\chi PT$ prediction \cite{Holstein:90},
$\alpha_E^{\pi^\pm} = (2.7 \pm 0.4)\times 10^{-4}\;fm^3$,
which is based on a robust relationship between
radiative pion decay $\pi\rightarrow e\nu\gamma$
and pion Compton scattering $\gamma\pi$ $\rightarrow$ $\gamma\pi$.
This discrepancy between theory and experiment obviously calls for more 
experimental attention.

Recently Gerasimov\cite{Gerasimov:98}, pointed out, that
the double radiative decay of the pionic deuterium 
($\pi^-d\rightarrow\gamma\gamma X$) can be 
used as a means to investigate the existence 
of the $d^*_1(1920)$ dibaryon ($\Gamma\sim 10$ keV) that was 
claimed by the DIB-2gamma collaboration\cite{Khrykin:97}. 
As pointed out by Gerasimov
\underline{if} the $d^*_1(1920)$ does indeed exist, one should
be able to detect it via the following chain:
\begin{center}
$\pi^-d\rightarrow d^*_1(1920)\gamma,\;\;\;d^*_1(1920)\rightarrow\gamma n n $
\end{center}

Based on a toy model, Gerasimov estimated that the resonant decay
of pionic deuterium via $d^*_1(1920)$ should occur roughly 100 times 
more frequently than non-resonant decay of $\pi^-p\rightarrow\gamma\gamma nn$.
Additionally, a clear signature of the hypothetical $d^*_1$ would be 
the photon spectrum with two lines: one at $\sim 90$ MeV 
($\pi^-d\rightarrow d^*_1\gamma$)
and the other  at $\sim 40$ ($d^*_1\rightarrow  \gamma n n $).

\section{TRIUMF $(\pi^-,\gamma\gamma)$ experiments and preliminary results}

The experiments presented here were performed at TRIUMF Laboratory 
using the RMC pair spectrometer~\cite{Wright:92} on the M9A beam line.
The pionic hydrogen data (E838) were taken in December 1998
and April-May 1999 and the pionic deuterium data (E864)
were taken in April 2000.

An 82 MeV/c pion beam was stopped in a liquid hydrogen (deuterium) target 
(approximate dimensions: length 15 cm, $\phi=$16cm),
The photons were converted into $e^+e^-$ pairs in a 1mm thick Pb cylinder.
The $e^+/e^-$ trajectories were then measured in a set of cylindrical wire and drift chambers,
and the magnetic field provided momentum analysis. Trigger scintillators
hit patterns were used to identify events.
\begin{figure}[hbt]
    \leavevmode
    \begin{center}
      \epsfig{figure=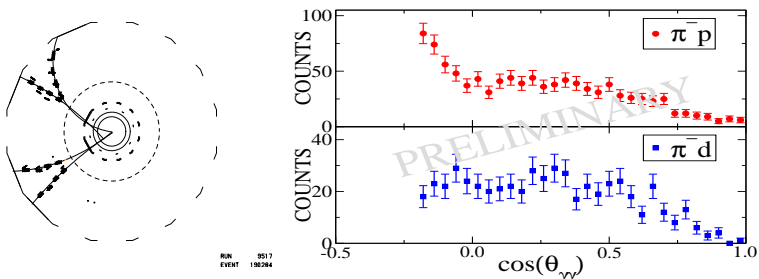,angle=0,width=10cm}
    \caption{A typical $\gamma\gamma$ event and opening angle distributions
	for pionic hydrogen (circles) and deuterium (squares).}
    \label{fig:rmc_results}
    \end{center}
\end{figure}

We collected approximately 1000 $\pi^-p\rightarrow\gamma\gamma n$ events 
and 500 $\pi^-d\rightarrow\gamma\gamma X$ events. 
A cut on the $\gamma\gamma$ opening angle
was used to remove $\pi^\circ\rightarrow\gamma\gamma$ events 
and a cut on the beam/trigger 
scintillators was used to remove accidental $\gamma\gamma$ events.

Fig.~\ref{fig:rmc_results} 
shows a typical event and the preliminary opening angle distributions 
for the pionic hydrogen and deuterium.
Our preliminary branching ratios are
B.R.$_{[\pi^-p\rightarrow\gamma\gamma n]}=3.8\times10^{-5}$ and 
B.R.$_{[\pi^-d\rightarrow\gamma\gamma X]}=1.6\times10^{-5}$.
The pionic hydrogen B.R. is in approximate agreement with Beder 
calculations~\cite{Beder:79} and the yield at small opening
angles demonstrates the dominance of the pion annihilation graph. 
We see no evidence for the the $d^*_1(1920)$ in our deuterium data.

In order to compare our results to the existing carbon 
data~\cite{DeutschMazzucato} we quote the relative 
branching ratios:
B.R.$_{\gamma\gamma/\gamma}=1.0\times10^{-4}$ for the pionic hydrogen and
B.R.$_{\gamma\gamma/\gamma}=0.6\times10^{-4}$ for the pionic deuterium.

\section{Summary}
Our preliminary results indicate that the 
the $\pi\pi\rightarrow \gamma\gamma$  graph is indeed dominant in 
the $\pi^- p \rightarrow \gamma \gamma n$ reaction hence this 
reaction is a potential probe for pion polarizability.
We have found no evidence for the $d^*_1(1920)$ in the decay 
of pionic deuterium.
We plan to extend our studies to nuclear targets.

% =========================================================================

\end{document}